\documentclass[prd,nofootinbib,floats,superscriptaddress,eqsecnum,tightenlines,11pt]{revtex4}
\usepackage{hyperref}
\usepackage{graphicx}
\usepackage{amsmath,amssymb,amsfonts,amsthm,latexsym}
\usepackage{marginnote}
\usepackage{color}
\usepackage{subfigure}

\def\be{\begin{equation}}
\def\ee{\end{equation}}
\def\beq{\begin{equation}}
\def\eeq{\end{equation}}
\newcommand{\bea}{\begin{eqnarray}}
\newcommand{\eea}{\end{eqnarray}}
\def\bi{\begin{itemize}}
\def\ei{\end{itemize}}
\def\ba{\begin{array}}
\def\ea{\end{array}}
\def\bfig{\begin{figure}}
\def\efig{\end{figure}}

\renewcommand\({\left(}
\renewcommand\){\right)}
\renewcommand\[{\left[}
\renewcommand\]{\right]}
\newcommand\del{\partial}

\newcommand{\ra}{\rightarrow}

\newcommand{\2}{{(2)}}
\newcommand{\3}{{(3)}}
\def\s{\sigma}

\newcommand{\aq}{a}   
\newcommand{\bb}{b} 

\def\ba{\begin{eqnarray}}
\def\ea{\end{eqnarray}}
\def\bas{\begin{subequations}\begin{eqnarray}}
\def\eas{\end{eqnarray}\end{subequations}}

\def\lp{\ell_\text{Pl}}

\def\la{\langle}
\def\ra{\rangle}
\def\de{\mathrm{d}}

\def\f{\frac}
\def\tf{\tfrac}

\def\su{\mathfrak{su}}

\def\mH{\mathcal{H}}
\def\mC{\mathcal{C}}
\def\mO{\mathcal{O}}
\def\nn{\nonumber}

\newcommand{\Dt}{{\cal D}_t}
\newcommand{\F}{{\cal F}}
\newcommand{\tg}{\tilde{g}}
\newcommand{\tX}{\tilde{X}}

\begin{document}

\title{Effective loop quantum cosmology as a higher-derivative scalar-tensor theory}

\author{David Langlois}
\email{langlois@apc.univ-paris7.fr}
\affiliation{Laboratoire Astroparticule et Cosmologie, Universit\'e Denis Diderot Paris 7, 75013 Paris, France}
\author{Hongguang Liu}
\email{hongguang.LIU@etu.univ-amu.fr}
\affiliation{Centre de Physique Th\'eorique (UMR CNRS 7332) , Universit\'es d'Aix-Marseille et de Toulon, 13288 Marseille, France}
\affiliation{Laboratoire de Math\'ematiques et Physique Th\'eorique (UMR CNRS 7350), Universit\'e Fran\c cois Rabelais, Parc de Grandmont, 37200 Tours, France}
\affiliation{Laboratoire Astroparticule et Cosmologie, Universit\'e Denis Diderot Paris 7, 75013 Paris, France}
\author{Karim Noui}
\email{karim.noui@lmpt.univ-tours.fr}
\affiliation{Laboratoire de Math\'ematiques et Physique Th\'eorique (UMR CNRS 7350), Universit\'e Fran\c cois Rabelais, Parc de Grandmont, 37200 Tours, France}
\affiliation{Laboratoire Astroparticule et Cosmologie, Universit\'e Denis Diderot Paris 7, 75013 Paris, France}
\author{Edward Wilson-Ewing}
\email{edward.wilson-ewing@unb.ca}
\affiliation{Max Planck Institute for Gravitational Physics (Albert Einstein Institute), Am M\"uhlenberg 1, 14476 Golm, Germany, EU}
\affiliation{Department of Mathematics and Statistics, University of New Brunswick, Fredericton, NB, Canada E3B 5A3}


\begin{abstract}

Recently, Chamseddine and Mukhanov introduced a higher-derivative scalar-tensor theory which leads to a modified Friedmann equation allowing for bouncing solutions. As we note in the present work, this Friedmann equation  turns out to reproduce exactly the loop quantum cosmology effective dynamics for a flat isotropic and homogeneous space-time.  We generalize this result to obtain a class of scalar-tensor theories, belonging to the family of mimetic gravity, which all reproduce the loop quantum cosmology effective dynamics for flat, closed and open isotropic and homogeneous space-times.

\end{abstract}

\maketitle

\section{Introduction}

Understanding the very early universe remains a fascinating open question in cosmology.  In the context of general relativity, an expanding universe containing ``standard'' matter fields  (which satisfy the null energy condition) is generically associated  with an initial singularity, where the space-time curvature becomes infinite. In this sense, classical general relativity fails to explain the very origin of our universe. 
When the value of the curvature approaches the Planck scale, quantum gravity effects are expected to become physically important  and  could  prevent the formation of space-time singularities. This is exactly what happens in the context of loop quantum cosmology (LQC) \cite{Ashtekar:2011ni, Bojowald:2008zzb, Banerjee:2011qu, Wilson-Ewing:2016yan} where quantum gravity effects are repulsive, in opposition to attractive classical gravity, and lead to a bouncing universe. 
 
However, quantizing gravity might not be necessary to resolve the initial cosmological singularity and one could envisage modifications of gravity at high curvature so that the singularity in general relativity is replaced by a bounce in a  modified gravity theory (see \cite{Battefeld:2014uga, Brandenberger:2016vhg} for recent reviews). Of course, these two approaches to avoid the singularity  could be two sides of the same coin if the classical equations derived from modified gravity can be interpreted as an effective description of the quantum behaviour.

Scalar-tensor theories provide a very large class of models for modified gravity theories. Among these, higher-order scalar-tensor theories, whose Lagrangians contain not only first order but also second order derivatives of the scalar field, have attracted a lot of attention lately. Allowing for higher-order time derivatives in the Lagrangian is potentially dangerous as this could lead to higher-order equations of motion which may require extra initial conditions and thus introduce an additional degree of freedom,  known as the Ostrogradsky ghost, because it leads to an Ostrogradsky instability \cite{Ostrogradsky:1850fid, Woodard:2015zca}. It is however  possible to find higher-order scalar-tensor theories that contain a single scalar degree of freedom (in addition to the usual tensorial modes associated with gravity) by imposing some restrictions on the initial Lagrangian. Initially, it was  believed  that a theory of this type was necessarily characterized by second order Euler-Lagrange equations, thus pointing to Horndeski theories \cite{Horndeski:1974wa} (see also \cite{Deffayet:2011gz}). In fact, requiring second order equations of motion turns out to be restrictive  and a much larger class of theories, dubbed  Degenerate Higher-Order Scalar-Tensor (DHOST) theories, has been recently identified, showing   that the absence of an extra unstable scalar mode is compatible with higher order Euler-Lagrange equations \cite{Gleyzes:2014dya, Gleyzes:2014qga, Langlois:2015cwa, Langlois:2015skt, Crisostomi:2016tcp, Crisostomi:2016czh, Achour:2016rkg, BenAchour:2016fzp}. These theories could provide an interesting arena to construct models for the early universe, as well as late-time cosmology.  Depending on whether corrections to general relativity appear at high-curvature scales and/or at large scales and low curvatures, the second order derivatives of the scalar field then correspond to ultraviolet and/or infrared corrections, and in particular high-curvature corrections can in some cases act as an ultraviolet cutoff like those that arise in a number of approaches to quantum gravity.

Among DHOST theories,  one can distinguish a special family of scalar-tensor theories that share properties similar to those of mimetic gravity.  Mimetic gravity is a higher order scalar-tensor theory which admits, in addition to the usual invariance under diffeomorphisms, a conformal invariance (which can be generalized to a conformal-disformal invariance).  Mimetic gravity was introduced in \cite{Chamseddine:2013kea} as a model for dark matter (see also \cite{Chamseddine:2014vna}). More recently, this model has been shown to admit (as a number of other scalar-tensor theories) non-singular bouncing cosmologies \cite{Chamseddine:2016uef, Ijjas:2016pad, Ijjas:2016tpn, Ijjas:2016vtq, Firouzjahi:2017txv}.

In the present work, we concentrate on  the specific mimetic gravity Lagrangian proposed in \cite{Chamseddine:2016uef}. We note that the corresponding classical equations of motion for a cosmological background are exactly the same as the so-called effective equations of loop quantum cosmology. This result suggests that it may be possible to describe loop quantum gravity at an effective level in some appropriate regimes as a higher-derivative scalar-tensor theory. The main purpose of this paper is to highlight this relation between mimetic gravity and loop quantum cosmology. While this relation has only been established at the cosmological level, it nonetheless provides a proposal for an effective description of loop quantum gravity in terms of higher-order scalar-tensor theories. Such an effective description is potentially very interesting, especially as it may give important insights into the relation between the quantization of gravity \`a la loop and the usual perturbative quantization techniques.

The paper is organized as follows. In the following section, we give a  short presentation of (degenerate) higher-order scalar-tensor theories and we present some basic properties of mimetic gravity, which can be seen as a particular example of these theories.  Then, in Sec.~\ref{sec:bounce}, we provide a brief review of loop quantum cosmology and explain how the effective equations are derived.  We then show how the loop quantum cosmology effective dynamics can be derived from an action principle $S[a,N]$ with a Lagrangian (invariant under time reparametrizations) that depends on the scale factor $a$ and on the lapse function $N$, for all isotropic cosmologies.  (This calculation is already known for the spatially flat case \cite{Date:2008gq, Helling:2009ia}.)  In Sec.~\ref{sec:lqg} we show that there exists a family of DHOST mimetic actions $S[g_{\mu\nu},\phi]$ which all reduce to $S[a,N]$ for homogeneous and isotropic space-time. These actions generalize the model proposed recently by Chamseddine and Mukhanov in \cite{Chamseddine:2016uef} and can be viewed as a proposal for an effective description of loop quantum gravity. We conclude in Sec.~\ref{sec:perspectives} with a discussion.

\section{Higher-Order Scalar-Tensor Theories}
\label{sec:DHOST}

In this section, we briefly review the main aspects of degenerate higher-order scalar-tensor (DHOST) theories. Their Lagrangian depends on a metric $g_{\mu\nu}$ and on a scalar field $\phi$, including its first and  second derivatives, $\nabla_\mu \phi \equiv \phi_\mu$ and $\nabla_\mu \!\nabla_\nu \phi \equiv \phi_{\mu\nu}$:
\bea\label{action}
S[g_{\mu\nu},\phi] \; = \; \int d^4x \, \sqrt{- g} \, {\cal L}(\phi,\phi_\mu,\phi_{\mu\nu}; g_{\mu\nu}) \, .
\eea
In general, such theories propagate an extra degree of freedom in addition to the usual scalar mode and the two tensor modes of the metric (assuming a linear dependence on the Riemann tensor%
\footnote{If the Lagrangian is not linear in  the Riemann tensor, then the theory can admit up to 8 degrees of freedom, most of them being unstable.}),
see \eqref{DHOST} below for an explicit example of such an action. This additional degree of freedom leads to instabilities (at least at the quantum level) and is known as an Ostrogradsky ghost \cite{Ostrogradsky:1850fid, Woodard:2015zca}. 

However, it is possible to find higher-order scalar-tensor theories that do not contain any Ostrogradsky ghost by imposing appropriate degeneracy conditions on the Lagrangian, thus defining DHOST theories. DHOST theories whose Lagrangian is at most cubic in $\phi_{\mu\nu}$ have already been classified \cite{BenAchour:2016fzp}.  In principle, this classification could be generalized to higher powers of $\phi_{\mu\nu}$.  

Below, we  first recall the main properties of DHOST theories and then concentrate on mimetic theories, which form a special family within DHOST theories.

\subsection{Evading the Ostrogradsky Instability}
\label{subsec:ostrogradsky}

Starting with a Lagrangian with second derivatives of $\phi$ is unusual in physics. Generically, the corresponding equation of motion for $\phi$ is fourth order in time derivatives, which means that more than two initial conditions (per space point) are required to fully specify the evolution. This signals the presence of an extra degree of freedom in the theory.  

However, there exist special Lagrangians with higher-order derivative terms for which the Euler-Lagrange equations remain second order. This is precisely the property verified by Horndeski theories in the context of scalar-tensor theories. It is even possible to find Lagrangians leading to third or fourth order Euler-Lagrange equations but without the dangerous extra  scalar mode. Examples of scalar-tensor theories of this type are the so-called beyond Horndeski theories, later encompassed in the DHOST theories. All these models are degenerate, a property which can also be seen in other contexts \cite{Motohashi:2016ftl, Klein:2016aiq, deRham:2016wji, Crisostomi:2017aim}.

By construction, DHOST theories satisfy some  degeneracy conditions so that they contain at most one scalar degree of freedom. To implement this degeneracy, it is useful to work  with a Hamiltonian formulation, based on the usual (3+1) ADM-decomposition of the metric on a space-time of the form $\Sigma \times \mathbb R$
\bea\label{ADM}
g_{\mu\nu}=
\left(
\begin{array}{cc}
-N^2+q_{ab}N^a N^b & q_{ab}N^b
\\
q_{ab}N^a & q_{ab}
\end{array}
\right)\,,
\eea
where $q_{ab}$ is the induced metric on the space slice $\Sigma$, $N$ is the lapse function and $N^a$ the shift vector. In this framework, the action \eqref{action} explicitly depends on second time derivatives of the scalar field and takes the general form\footnote{We do not consider theories which involve, after a (3+1)-decomposition, second time derivatives of the metric components which are not total derivatives. Such theories are expected to propagate Ostrograsky ghosts that cannot be removed.} (up to boundary terms that we neglect)
\bea\label{3+1action}
S[g_{\mu\nu},\phi] \; = \; \int d^4x \, N \sqrt{q} \, {\cal L}(q_{ab},K_{ab},N,N^a;\phi,A_*,\dot{A}_*) \, ,
\eea
with 
\bea\label{KA}
K_{ab} \equiv \frac{1}{2N} \left( \dot{q}_{ab} - D_a N_b - D_b N_a \right) \,, \qquad
A_* \equiv \frac{\dot\phi- N^a \partial_a \phi}{N} \, ,
\eea
where $D_a$ is the covariant derivative compatible with $q_{ab}$. For simplicity, we use the same notation for the Lagrangian densities $\cal L$ in the covariant \eqref{action} and the non-covariant \eqref{3+1action} versions of the action, even though they are not strictly speaking the same function. 
 
To perform the Hamiltonian analysis \cite{Langlois:2015skt}, it is convenient to use the auxiliary variable $A_*$ as an independent variable, so that all second time derivatives of $\phi$ are absorbed in $\dot A_*$. This procedure thus introduces a new pair of variables, $A_*$ and its conjugate momentum, which a priori describes an extra scalar degree of freedom. However, it is still possible that the theory propagates no more than one scalar degree of freedom if there exist constraints (in addition to the usual four constraints associated with space-time diffeomorphism invariance) so that the effective number of physical degrees of freedom is reduced. The existence of a primary constraint is equivalent to the requirement that the Hessian matrix of \eqref{3+1action} (whose coefficients are the second derivatives of the action with respect to velocities of the fields) is degenerate. This property of degeneracy of the Hessian matrix has been used systematically to construct DHOST Lagrangians, initially with a quadratic dependence on $\phi_{\mu\nu}$ \cite{Langlois:2015cwa} and, more recently, with a cubic dependence \cite{BenAchour:2016fzp}. Note that the primary constraint is usually of the second-class type and imposing its time conservation  leads to a secondary constraint, which is also second-class. Both constraints thus eliminate the dangerous extra degree of freedom \cite{Langlois:2015skt}. The special case where the primary constraint is first-class, signalling an additional local symmetry of the action, is seen in the mimetic models, which will be discussed in the next subsection.  

All the  DHOST theories that have been identified can be written in the form 
\bea
\label{DHOST}
S[g,\phi] &=& \int d^4 x \, \sqrt{- g }
\left[ {f_2(X,\phi)} \,  R +  C_\2^{\mu\nu\rho\sigma} \,  \phi_{\mu\nu} \, \phi_{\rho\sigma}
\right.
\cr
&&
\left. \qquad \qquad\qquad 
+ f_3(X, \phi) \, G_{\mu\nu} \phi^{\mu\nu}  +  
C_\3^{\mu\nu\rho\sigma\alpha\beta} \, \phi_{\mu\nu} \, \phi_{\rho\sigma} \, \phi_{\alpha \beta} \right]  \;,
\eea
where the functions $f_2$ and $f_3$ depend only on the scalars $\phi$ and $X \equiv \phi_\mu \phi^\mu$;  $R$ and $G_{\mu\nu}$ denote, respectively, the usual Ricci scalar and Einstein tensor associated with the metric $g_{\mu\nu}$. The tensors  $C_\2$ and $C_\3$ are the most general tensors constructed from the metric $g_{\mu\nu}$ and the first derivative of the scalar field $\phi_\mu$. It is easy to see that the quadratic terms can be rewritten as
\beq
\label{C2}
C_\2^{\mu\nu\rho\sigma} \,  \phi_{\mu\nu} \, \phi_{\rho\sigma} =\sum_{A=1}^{5}a_A(X,\phi)\,   L^\2_ A\,,
\eeq
with the elementary quadratic Lagrangians (i.e., terms quadratic in $\phi_{\mu\nu}$ or $\Box\phi$; since $\phi_\mu$ terms only contain one derivative, they do not contribute to the order of DHOST terms) given by
\be
\label{QuadraticL}
\begin{split}
& L^\2_1 = \phi_{\mu \nu} \phi^{\mu \nu} \,, \qquad
L^\2_2 =(\Box \phi)^2 \,, \qquad
L_3^\2 = (\Box \phi) \phi^{\mu} \phi_{\mu \nu} \phi^{\nu} \,,  \\
& L^\2_4 =\phi^{\mu} \phi_{\mu \rho} \phi^{\rho \nu} \phi_{\nu} \,, \qquad
L^\2_5= (\phi^{\mu} \phi_{\mu \nu} \phi^{\nu})^2\,.
\end{split}
\ee
In a similar fashion, the cubic terms can be written as
\beq
\label{C3}
C_\3^{\mu\nu\rho\sigma\alpha\beta} \, \phi_{\mu\nu} \, \phi_{\rho\sigma} \, \phi_{\alpha \beta} = \sum_{A=1}^{10} b_A(X,\phi)\,  L^\3_A \,,
\eeq
with the elementary cubic Lagrangians being
\be
\label{CubicL}
\begin{split}
& L^\3_1=  (\Box \phi)^3  \,, \quad
L^\3_2 =  (\Box \phi)\, \phi_{\mu \nu} \phi^{\mu \nu} \,, \quad
L^\3_3= \phi_{\mu \nu}\phi^{\nu \rho} \phi^{\mu}_{\rho} \,,   \\
& L^\3_4= \left(\Box \phi\right)^2 \phi_{\mu} \phi^{\mu \nu} \phi_{\nu} \,, \quad
L^\3_5 =  \Box \phi\, \phi_{\mu}  \phi^{\mu \nu} \phi_{\nu \rho} \phi^{\rho} \,, \quad
L^\3_6 = \phi_{\mu \nu} \phi^{\mu \nu} \phi_{\rho} \phi^{\rho \s} \phi_{\s} \,,   \\
& L^\3_7 = \phi_{\mu} \phi^{\mu \nu} \phi_{\nu \rho} \phi^{\rho \s} \phi_{\s} \,, \quad
L^\3_8 = \phi_{\mu}  \phi^{\mu \nu} \phi_{\nu \rho} \phi^{\rho}\, \phi_{\s} \phi^{\s \lambda} \phi_{\lambda} \,,   \\
& L^\3_9 = \Box \phi \left(\phi_{\mu} \phi^{\mu \nu} \phi_{\nu}\right)^2  \,, \quad
L^\3_{10} = \left(\phi_{\mu} \phi^{\mu \nu} \phi_{\nu}\right)^3 \,.
\end{split}
\ee
In general, as explained above, theories with an action of the form \eqref{action} contain two tensor modes and two scalar modes, one of which leads to an Ostrogradsky instability. DHOST theories correspond to specific restrictions of the functions $\aq_A$ and $\bb_A$, so that the  Hessian matrix is degenerate. One finds that these theories contain at most one propagating scalar mode. DHOST theories include Horndeski and beyond-Horndeski theories but many other higher-order scalar-tensor theories as well.

\subsection{Mimetic theories}
\label{subsec:mim}

Originally, mimetic gravity was introduced by Chamseddine and Mukhanov as a scalar-tensor theory defined by the usual Einstein-Hilbert action \cite{Chamseddine:2013kea}
\bea\label{actionCM}
S_{CM}[\tg_{\mu\nu},\phi] \; = \; S_{EH}[{g}_{\mu\nu}] \, \equiv \, 
\frac{1}{16\pi G} \int d^4 x \, \sqrt{- {g}}  \,  R[{g}_{\mu\nu}] 
\eea
where the metric ${g}_{\mu\nu}$ is related to $\tg_{\mu\nu}$ and $\phi$ by the non-invertible conformal transformation
\bea
{g}_{\mu\nu} \; \equiv \; -\tX \, \tg_{\mu\nu} \, \qquad \text{with} \qquad \tX \; \equiv  \; \tg^{\mu\nu} \phi_\mu \phi_\nu \, .
\eea 
It is immediate to see that the metric ${g}_{\mu\nu}$ is invariant under a local conformal transformation of the metric $\tg_{\mu\nu}$ while $\phi$ is left unchanged
\bea \label{conformal-sym}
\tg_{\mu\nu} \longmapsto \Omega(x) \, \tg_{\mu\nu} \, , \qquad \phi \longmapsto \phi \, .
\eea 	
Here $\Omega$ is an arbitrary function on the space-time. Hence, the action \eqref{actionCM} is invariant under this same local transformation provided that the coupling to matter (if there is any) is defined with respect to ${g}_{\mu\nu}$.

There exist two different equivalent reformulations of the mimetic action \eqref{actionCM} as it was emphasized in \cite{Arroja:2015wpa}. 
The observation that
\bea\label{mimcondition}
{X} \equiv {g}^{\mu\nu} \phi_\mu \phi_\nu \, = \, 
-\frac{1}{\tX} \tg^{\mu\nu} \phi_\mu \phi_\nu \; = \; -1 \, ,
\eea
enables us to replace the action \eqref{actionCM} by the following one
\bea\label{actionCMl}
S_{CM}^{(1)}[{g}_{\mu\nu},\phi;\lambda] \; \equiv \;  S_{EH}[{g}_{\mu\nu}]  + \int d^4 x \, \sqrt{-{g}} \, \lambda ({X} + 1) \, ,
\eea
where the equation of motion for $\lambda$ reproduces exactly the condition \eqref{mimcondition}. It follows that the actions \eqref{actionCM}  and \eqref{actionCMl} are classically equivalent.

In the second reformulation of mimetic gravity, one considers the action written in terms of the metric $\tg_{\mu\nu}$ and $\phi$. 
Using the transformation law of the Ricci tensor  under a conformal transformation of the metric, the action \eqref{actionCM} can be rewritten as the following higher-derivative scalar-tensor theory
\bea\label{actionCMII}
S_{CM}^{(2)}[\tg_{\mu\nu},\phi] \; \equiv \; \frac{1}{16\pi G}\int d^4 x \, \sqrt{-\tg}  \, \left( \tX R[\tg_{\mu\nu}] + \frac{6}{\tX} \tilde{\phi}_{\mu}^{\  \nu}  \tilde{\phi}^{\mu \rho} \phi_\nu \phi_\rho\right) \, ,
\eea
where $\tilde{\phi}_{\mu\nu} = \tilde{\nabla}_\mu \phi_\nu$ with $\tilde{\nabla}$ being the covariant derivative compatible with $\tg_{\mu\nu}$, and indices are lowered and raised with $\tg_{\mu\nu}$ and its inverse.
The action \eqref{actionCMII} is clearly of the form \eqref{C2}, where the only non-trivial
coefficients are
\bea
f_2 \, = \, \frac{\tX}{16\pi G} \quad \text{and} \quad a_4 \; = \; {\frac{3}{8 \pi G \tX}} \, .
\eea
As expected, this theory is degenerate and can be shown to belong to the class Ia (according to the classification given in \cite{Achour:2016rkg}) which corresponds to Lagrangians which are obtained from a conformal-disformal transformation of the quadratic Horndeski action. This is in total agreement with the fact that the mimetic gravity action has been obtained from a (non-invertible) conformal transformation of the Einstein-Hilbert action (which can be viewed as a particular case of Horndeski theory).  A Hamiltonian analysis of the mimetic action \eqref{actionCMl} (and generalizations thereof) can be found in \cite{Kluson:2017iem}.

We close this short review with a remark concerning generalizations of mimetic gravity. First of all, one can generalize the action \eqref{actionCM} assuming that ${g}_{\mu\nu}$ is now a general non-invertible conformal-disformal transformation of $\tg_{\mu\nu}$ \cite{Deruelle:2014zza}, i.e.,
\bea\label{generaltildeg}
{g}_{\mu\nu} \; \equiv \; A(\tX,\phi) \tg_{\mu\nu} + B(\tX,\phi) \phi_\mu \phi_\nu \quad \text{with} \quad \frac{\partial}{\partial \tX}(A+ \tX B) = 0 \, .
\eea
In that case, it is easy to see that the local conformal invariance of ${\tg}_{\mu\nu}$ has been generalized to invariance under the local symmetry
\bea\label{gensym}
\delta \tg_{\mu\nu} = \alpha(x) \tg_{\mu\nu} + \beta(x) \phi_\mu \phi_\nu \quad \text{with} \quad
(A-\tilde{X}A_{\tilde{X}}) \alpha(x) = \tilde{X}{}^2 A_{\tilde{X}} \beta(x) \,,
\eea
where $\alpha(x)$ and $\beta(x)$ are functions on the space-time and $A_{\tilde{X}} \equiv \del_{\tilde{X}} A$. Hence, we obtain in this way an action $S[\tg_{\mu\nu},\phi]$ that is invariant under the symmetry \eqref{gensym}, and therefore the theory is degenerate. In fact, given any higher-order scalar-tensor action ${S}[{g}_{\mu\nu},\phi]$, the action defined by
\bea
\tilde{S}[\tg_{\mu\nu},\phi] \equiv {S}[{g}_{\mu\nu},\phi]
\eea
where ${g}_{\mu\nu}$ is defined by \eqref{generaltildeg} is necessarily degenerate. This family of actions provides a large generalization of mimetic gravity theories.

\section{Loop Quantum Cosmology and its Effective Dynamics}
\label{sec:bounce}

As discussed in the introduction, the recent work \cite{Chamseddine:2016uef} by Chamseddine and Mukhanov, based on a particular model of mimetic gravity, leads to a modified Friedmann equation that coincides with the effective dynamics of loop quantum cosmology (LQC). This connection was not mentioned in  \cite{Chamseddine:2016uef}.
 In order to clarify this link, we first briefly review LQC for spatially flat FLRW space-times and  then perform a Hamiltonian analysis for a large class of mimetic gravity theories restricted to (homogeneous and isotropic) cosmological geometries. This allows us to dermine the conditions on the mimetic gravity action under which the effective dynamics of LQC are recovered.  Interestingly, we find that there exists a large class of such theories, among which the Chamseddine-Mukhanov model is a particular example.  Furthermore, we extend this analysis  to the case of spatially curved FLRW space-times and find, once again, that there exists a large class of higher-derivative scalar-tensor theories which give the LQC effective dynamics for all FLRW space-times.

\subsection{Loop Quantum Cosmology: A Review}

LQC is a proposal for quantizing cosmological space-times using the variables and the non-perturbative techniques of LQG.  More specifically, as in the full theory, LQC is based on the Ashtekar-Barbero connection variables, and the elementary variables to be promoted to fundamental operators in the quantum theory are holonomies along edges and fluxes across surfaces.  For dedicated reviews of LQC, see, e.g., \cite{Ashtekar:2011ni, Bojowald:2008zzb, Banerjee:2011qu, Wilson-Ewing:2016yan}.

\subsubsection{Hamiltonian Framework with Ashtekar-Barbero Variables}

The Ashtekar-Barbero variables are related to the metric and extrinsic curvature as follows. We first introduce the densitized triads $E^a_i = \sqrt{q} e^a_i$ (where $q$ is the determinant of the spatial metric and the triads $e^a_i$ are related to the inverse of the spatial metric by $q^{ab} = e^a_i e^b_j \delta^{ij}$).
The conjugate variable to $E^a_i$ is the $\su(2)$-valued Ashtekar-Barbero connection 
\bea
A_a^i = \Gamma_a^i + \gamma K_a^i \, , 
\eea
where $K_a^i = K_{ab} e^b_i$ encodes the extrinsic curvature, $\Gamma_a^i$ is the spin-connection such that 
\bea
D_a e^b_i \,\equiv \, \del_a e^b_i - \Gamma^b_{ac} e^c_i + \epsilon_{ij}{}^k \Gamma_a^j e^b_k \, = \, 0 \; ,
\eea
with $\Gamma^b_{ac}$ being the usual Christoffel symbols on the spatial slice, and $\gamma$ is the real-valued Barbero-Immirzi parameter. The symplectic structure of gravity in the Ashtekar-Barbero variables is given by the Poisson bracket
\bea\label{PoissonAE}
\{ A_a^i(x) \, , \, E^b_j(y)\} \; = \; {8\pi G \gamma} \, \delta_a^b \, \delta_j^i \, \delta^{(3)}(x-y).
\eea
For simplicity, the matter field is often assumed to be a scalar $\psi$, with a Lagrangian
\be
L_\psi = \int \de^4x \, \sqrt{-g} \[ -\tf{1}{2} (\del \psi)^2 - V(\psi) \].
\ee
In the Hamiltonian framework, $\psi$ comes with a conjugate momentum $\pi_\psi$ such that
\bea
\{\psi, \pi_\psi\} = 1 \,  \qquad \text{and} \qquad \pi_\psi \equiv \frac{\partial L_\psi}{\partial \dot{\psi}}= \frac{\sqrt{q}}{N} \, \dot{\psi} \, .
\eea
The lapse function $N$ in the metric is not a dynamical variable and, in the Hamiltonian formulation, plays the role of a Lagrange multiplier that enforces that the scalar constraint $\mH$ vanish, with $\mH$ given by \cite{Barbero:1994ap, Holst:1995pc, Thiemann:2007zz}
\be \label{gen-ham}
\mH = -\f{E^a_i E^b_j}{16 \pi G \gamma^2 \sqrt{q}} \, \epsilon^{ij}{}_k \(F_{ab}{}^k - (1 + \gamma^2) \Omega_{ab}{}^k \) + \f{\pi_\psi^2}{2 \sqrt{q}} + \sqrt{q} \, V(\psi),
\ee
where $F_{ab}{}^k = 2 \del_{[a} A_{b]}^k + \epsilon_{ij}{}^k A_a^i A_b^j$ is the field strength of the Ashtekar-Barbero connection, while the tensor $\Omega_{ab}{}^k = 2 \del_{[a} \Gamma_{b]}^k + \epsilon_{ij}{}^k \Gamma_a^i \Gamma_b^j$ measures the spatial curvature.

In the particular case of a spatially flat FLRW space-time, with metric
\be\label{FLRWmetric}
\de s^2 = -N^2 \de t^2 + a(t)^2 \de \vec x^2\,,
\ee
the densitized triads $E^a_i$ are given by
\be \label{def-p}
E^a_i = p \( \f{\del}{\del x^i} \)^a\, \qquad \text{with} \qquad p = a^2\,,
\ee
and the Ashtekar-Barbero connection can be written as
\be \label{def-c}
A_a^i = c \, (\de x^i)_a\, \qquad \text{with} \qquad c =\frac{\gamma \,  \dot{a}}{N}\,,
\ee
since $\Gamma_a^i=0$ because of the space-time symmetries. Therefore, the gravitational sector of the phase space is two-dimensional, and the symplectic structure, inherited from \eqref{PoissonAE}, reduces to\footnote{As a technical remark, note that for the symplectic structure to be well-defined in such a space-time, it is necessary to restrict integrals to some finite region, called the fiducial cell.  This restriction on the integrals acts as an infrared regulator which should be removed once the dynamics are determined by taking the limit of the fiducial cell going to the entire spatial manifold.  For simplicity, here we choose the fiducial cell such that its volume with respect to the metric $\de \mathring{s}^2 = \de \vec x^2$ is 1.} 
\be
\{c, p\} = \f{8 \pi G \gamma}{3}.
\ee
Moreover, the scalar constraint (\ref{gen-ham}) becomes
\be \label{flrw-ham}
\mH = -\f{3}{8 \pi G \gamma^2} {p}^{1/2} \, c^2 + \f{\pi_\psi^2}{2 p^{3/2}} + p^{3/2} V(\psi).
\ee
Note that $\Omega_{ab}{}^k = 0$ for the spatially flat FLRW space-time. The dynamical evolution of any observable $\mO$ is then given by the smeared Hamiltonian constraint according to
\be
\label{evol}
\f{\de \mO}{\de t} = \{ \mO, \mC_H\}\, \qquad \text{with} \qquad
\mC_H = \int \de^3 \vec x \, N \mH \, . 
\ee
Note that the spatial diffeomorphism and Gauss constraints cannot contribute to the smeared Hamiltonian constraint since they identically vanish for the choice of variables \eqref{def-p} and \eqref{def-c}. Also, the spatial integral of $N \mH$ is trivial in FLRW space-times since every term is independent of position due to homogeneity. One can use the relation \eqref{evol} with $\mO=p$ to recover the usual Friedmann equation.

\subsubsection{Quantum Theory}
\label{ssubsec:lqc-q}

The quantization of such a theory (assuming now for simplicity that $V(\psi) = 0$) following the standard Wheeler-de Witt procedure will give a quantum cosmology where the classical big-bang singularity is not resolved in any meaningful sense.  Indeed, sharply-peaked wave packets closely follow the classical (singular) solutions, and the expectation value of, e.g., the energy density of $\psi$ can become arbitrarily large \cite{Ashtekar:2011ni}.

The situation is markedly different in LQC for the reason that the fundamental operators of the theory are holonomies and areas, not operators corresponding to the connection $A_a^i$ itself.  For this reason, in LQC it is not possible to directly promote the symmetry-reduced scalar constraint \eqref{flrw-ham} to an operator in the quantum theory since there is no operator corresponding to $\hat c$.  Instead, it is necessary to go back one step and construct an operator corresponding to \eqref{gen-ham}.  

This can be done in two parts.  First, since the $p$ contained in the $E^a_i$ corresponds to an area, it can directly be promoted to be an operator.  Second, the $\su(2)$-valued field strength can be expressed in terms of holonomies in the same fashion as in lattice gauge theories,
\be \label{f-hol}
F_{ab} \simeq \f{h_{\Box_{ab}} - \mathbb{I}}{Ar_\Box},
\ee
where $h_{\Box_{ab}}$ is the holonomy of the connection $A_a$ around a loop in the $a$--$b$ plane, $\mathbb I$
is the identity and $Ar_\Box$ is the area of that loop.  In a lattice gauge theory, one would be interested in the limit of the right-handside of \eqref{f-hol} when $Ar_\Box \to 0$, in which case the relation \eqref{f-hol} becomes exact.  However, this is not natural in LQC, since the spectrum of the area operator in LQG is discrete and has a minimum non-zero eigenvalue
\bea\label{delta}
\Delta = 4 \sqrt 3 \pi \gamma \lp^2
\eea
where $ \lp$ is the Planck length.  Therefore, what is done in LQC is to express $F_{ab}$ in terms of the holonomy of $A_a$ around a loop with this minimal area $\Delta$.

More specifically, given the symmetries of the FLRW space-time, the loop $\Box_{ab}$ is assumed to be a square loop in the $a$--$b$ plane.  The holonomy of $A_a$ along edges parallel transported by the vectors $(\del / \del x^i)^a$ is easily evaluated.  Since the Ashtekar-Barbero connection $A_a = A_a^i \tau_i$ is $\su(2)$-valued%
\footnote{\label{footnote-tau}
The elements $\tau_i$ form a basis of the $\su(2)$ Lie algebra satisfying
\bea
[\tau_i, \tau_j] = 2 \epsilon_{ij}{}^k \tau_k  \quad \text{and} \quad  {\rm Tr} (\tau_i \tau_j) = - {2} \delta_{ij} \, . \nonumber
\eea
We used the notation $\epsilon_{ij}{}^k$ for the totally antisymmetric symbol with $\epsilon_{123}=+1$, indices are raised and lowered with $\delta_{ij}$, and $\text{Tr}$ denotes the trace in the 2-dimensional fundamental representation (known as the Killing form). In the spin-$1/2$ representation, the elements $\tau_i$ are represented by the 2-dimensional matrices
\bea
\tau_1=\left( \begin{array}{cc} 0 & - i \\ -i & 0 \end{array}\right)\, , \quad
\tau_2=\left( \begin{array}{cc} 0 & - 1 \\ 1 & 0 \end{array}\right)\, , \quad 
\tau_3=\left( \begin{array}{cc} -i & 0 \\ 0 & i \end{array}\right)\, . \nonumber
\eea
},
it is necessary to choose a representation in which to calculate the holonomy.  This is usually chosen to be the $j=1/2$ representation. This is not only the simplest non-trivial representation, but also corresponds to the smallest excitation $\Delta$ of area possible in LQG, which is precisely the area that has been chosen for the loop $\Box_{ab}$ to have from physical grounds, as argued in the previous paragraph.

In the $j=1/2$ representation, the $\tau_i$ can be chosen to be the Pauli matrices (up to a factor of $i/2$ in order to have the correct normalization as shown in footnote \ref{footnote-tau}). Hence, the holonomy of $A_a$ along a path parallel to $(\del / \del x^i)^a$ and of length $\ell$ with respect to the fiducial metric $\de \mathring{s}^2 = \de \vec x^2$, is (no sum over $i$)
\be \label{hi}
h_i(\ell) = \exp \( \int_0^\ell \de x^i A_a \( \f{\del}{\del x^i} \)^a \) = \cos \f{\ell c}{2} \mathbb{I} + 2 \sin \f{\ell c}{2} \tau_i \, .
\ee
Note that the fiducial metric allows to identify internal (Lie algebra) indices $i,j,k,\cdots$ with space indices $a,b,c,\cdots$ so that, from now on, we will use the same notation $i,j,k,\cdots$ to label indifferently internal and space directions. An important point here is that the length $\ell$ is measured with respect to the fiducial metric, and the physical length of the edge along which $h_i$ is evaluated is given by $a \, \ell$ where $a$ is the scale factor.  Therefore, requiring that the physical length of $h_i$ be $\sqrt{\Delta}$ corresponds to setting $\sqrt{p} \, \ell = \sqrt\Delta \Rightarrow \ell = \sqrt{\Delta/p}$ where we used $a=\sqrt{p}$. Then, the holonomy $h_{\Box_{ij}}$ around a square loop in the $x^i$--$x^j$ plane with a physical area equal to $\Delta$ \eqref{delta} 
is
\be \label{hBox}
h_{\Box_{ij}} = h_j(\bar{\mu})^{-1} h_i(\bar{\mu})^{-1} h_j(\bar{\mu}) h_i(\bar{\mu}),
\ee
where $\bar\mu = \sqrt{\Delta / p}$, and $Ar_\Box = \bar\mu^2$.

Now, the holonomy \eqref{hBox} can be defined in the quantum theory, and therefore can be used as the operator corresponding to $\hat F_{ab}$.  Then, given the field strength operator, it is now easy to define operators corresponding to the scalar constraint, and this completes the quantum theory.  For the precise details concerning the Hilbert space and the Hamiltonian constraint operator (which are not necessary here for our purposes), see, e.g., \cite{Ashtekar:2011ni, Banerjee:2011qu}.  

The resulting quantum theory resolves the big-bang singularity in a precise sense: first, there is an upper bound on the operator corresponding to the matter energy density, and second, the states corresponding to singular space-times (i.e., $p = a^2 = 0$) decouple from non-singular states under the action of the Hamiltonian constraint operator and thus an initial state which is non-singular will always remain non-singular.  These important differences from the Wheeler-de Witt theory arise from expressing the field strength operator in terms of the holonomy of the connection around a loop of area $\Delta$ rather than directly promoting $c$ to be an operator.

Furthermore, for the case that the scalar field $\psi$ is massless, it provides a good relational clock and it is possible to speak of the relational evolution of the wave function $\Psi(p, \psi)$ with respect to $\psi$.  In particular, it is possible to construct an `initial' state $\Psi(p, \psi_o)$ at some `initial' relational time $\psi_o$ and to evolve it using the Hamiltonian constraint operator.  One especially interesting possibility is to construct a wave packet sharply peaked around a classical configuration at a low curvature scale when general relativity can be trusted, and then evolve the wave packet towards the high-curvature regime.  An important result in this case is that the wave packet remains sharply peaked throughout its entire evolution, assuming (i) it is initially sharply peaked and (ii) that the expectation value of $p$ always remains large compared to $\lp^2$.  For such states, the full quantum dynamics are extremely well approximated by an effective theory \cite{Ashtekar:2006wn, Taveras:2008ke, Rovelli:2013zaa}.

\subsubsection{Effective Dynamics}
\label{ssubsec:effective}

The effective dynamics of LQC are obtained by expressing the classical field strength $F_{ab}$ in terms of the holonomy around a square loop of area $\Delta$ as we did in the previous section, and then treating the resulting $\mH$ classically.  It is easy to verify that the LQC effective scalar constraint is 
\be \label{eff-ham}
\mH^{\rm eff} = -\f{3 \, p^{3/2}}{8 \pi G \Delta \gamma^2} \, \sin^2 \bar\mu c + \f{\pi_\psi^2}{2 p^{3/2}} + p^{3/2} V(\psi).
\ee
In this way, the effective theory captures the physics corresponding to the discrete nature of geometry in LQC (specifically, the existence of the area gap $\Delta$), but ignores the effect of quantum fluctuations since $\mH^{\rm eff}$ is treated classically\footnote{{This is a good approximation for states that are initially sharply peaked (i.e., all expectation values satisfy $\la \hat \mO^2 \ra \approx \la \hat \mO \ra^2$ which means that quantum fluctuations are negligeable) so long as the spatial volume of the space-time is much larger than the Planck volume since for these states quantum fluctuations do not grow significantly and hence always remain negligeable \cite{Rovelli:2013zaa}.}}.

Using \eqref{evol} with \eqref{eff-ham}, it is easy to derive the time derivative of $p$:
\be
\dot p=2N \frac{p}{\gamma\sqrt{\Delta}} \sin \bar\mu c \cos \bar\mu c\,.
\ee
Moreover, the constraint \eqref{eff-ham} implies 
\beq
\sin^2 \bar\mu c=\frac{\rho}{\rho_c},
\eeq
where $\rho$ is the matter energy density, 
\beq
\rho = \f{\pi_\psi^2}{2 p^{3}} +  V(\psi)\,,
\eeq
and $\rho_c$ is a constant defined by
\beq\label{rhoc}
\rho_c = \frac{3}{8\pi G \gamma^2 \Delta}\,= \, \frac{\sqrt{3}}{32\pi^2 \gamma^3} \rho_{\rm Pl} \, ,
\eeq
with $\rho_{\rm Pl}$ being the Planck density. Combining the above relations yields the effective Friedmann equation
\be \label{lqc-eff}
H^2 = \( \f{\dot p}{2 N p} \)^2 = \f{8 \pi G}{3} \rho \(1 - \f{\rho}{\rho_c} \),
\ee
where $H = \dot a /(N a)$ is the Hubble parameter. It can also be checked that the continuity equation in effective LQC is the same as in classical general relativity.

As can easily be seen from the LQC effective Friedmann equation, the big-bang singularity of general relativity is replaced by a bounce that occurs when the energy density of $\psi$ reaches the critical energy density $\rho_c \propto \rho_{\rm Pl}$ \eqref{rhoc}.  This bounce clearly originates from the quantum geometry of LQG, and in the limit of $\Delta \to 0$, the classical Friedmann equation is recovered.  Numerical simulations of the dynamics generated by the LQC Hamiltonian constraint operator  for sharply-peaked wave functions have explicitly shown that the effective equations do indeed provide an excellent approximation to the full quantum dynamics, even around and at the bounce point \cite{Ashtekar:2006wn}.  Thus, it is clear that the bounce occurs due to the discrete geometry of LQG, irrespective of quantum fluctuations.

Note that the bounce occurs when the energy density is of the order of the Planck scale but the volume of the spatial slice at the bounce time can be much larger than the Planck volume.  Consequently, as long as the bounce occurs at a spatial volume much larger than $\lp^3$,  the effective theory can be trusted at all times for sharply-peaked states. Note that this condition is automatically satisfied for non-compact FLRW space-times since their spatial volume is always infinite (so long as the scale factor remains non-vanishing, which is always true in LQC).

In conclusion, the LQC effective dynamics is a powerful tool which significantly simplifies calculations of quantum gravity effects in semi-classical cosmological states.  Clearly, it would be very helpful if it were possible to develop an effective theory that would hold more generally, for instance for all states in LQG that have nice semi-classical properties and whose geometric observables of interest concern (spatial) regions that are much larger than $\lp^3$.  For this reason, we explore  scalar-tensor theories that are able to reproduce the LQC effective dynamics for cosmological geometries.

\subsection{Loop Quantum Cosmology from Mimetic Gravity}

The goal now is to find a family of modified gravity theories that, when restricted to the spatially flat FLRW space-time, reproduce precisely the LQC effective Friedmann equation \eqref{lqc-eff}. In fact, one such modified gravity theory with precisely this property has already been found \cite{Date:2008gq, Helling:2009ia} (see also \cite{Olmo:2008nf} for an $f(R)$ modified gravity theory whose dynamics are a good approximation to the LQC Friedmann equation). Here, we will generalize these earlier results to a whole class of scalar-tensor theories, and in Sec.~\ref{subsec:curv} we extend these results to the case of non-vanishing spatial curvature.

To begin, we look for an action $S[a,N,\psi]$ where the dynamical variables are the scale factor $a(t)$, the lapse function $N(t)$ and a field $\psi(t)$ that represents the matter content of the universe.  The action is of course invariant under time reparametrizations. Afterwards, we will construct a class of covariant actions of modified gravity which reduce to $S[a,N,\psi]$ when the metric is fixed by the flat FLRW metric \eqref{FLRWmetric}. We assume that the field $\psi$ is a massless scalar field minimally coupled to gravity. Hence, the modified action of gravity we are looking for takes the form
\bea
\int d^4x \, \sqrt{\vert g \vert} \, \left( \frac{1}{16\pi G} R - \frac{1}{2}g^{\mu\nu} \psi_\mu \psi_\nu  + \cdots \right) \, ,
\eea
where the remaining (so far unknown) part does not involve the matter content represented here by $\psi$. As a consequence, on an FLRW space-time, the previous action reduces to
\bea \label{FLRWaction}
S[a,N,\psi] \; = \; \int dt \left( - \frac{3a \dot{a}^2}{8 \pi G N} + a^3 \frac{\dot{\psi}^2}{2 N} + Na^3 {\cal L}\(a,\frac{\dot{a}}{N}\)\right)
\eea
where the unknown function $\cal L$ has to be fixed in such a way that $S[a,N,\psi]$ reproduces the LQC effective dynamics. The fact that $\cal L$ depends on $\dot{a}/N$ (rather than on $\dot{a}$ and $N$ separately) is a consequence of requiring invariance under time reparametrization. Furthermore, the Lagrangian does not involve non-trivial higher derivatives (which cannot be eliminated from the action with integrations by parts) of the scale factor, otherwise the associated classical equations of motion would (necessarily) be higher order, hence they would not reproduce the LQC effective dynamics. 

As can be seen from \eqref{eff-ham}, up to an overall prefactor the effective Hamiltonian constraint of loop quantum cosmology is expressed only in terms of the combination $c / \sqrt{p}$, which classically corresponds to the Hubble rate $H = {\dot{a}}/{(Na)} $ (up to the prefactor $\gamma \sqrt\Delta$), as can be seen from the definitions \eqref{def-p} and \eqref{def-c}.  This suggests restricting the function $\cal L$ to be to the form
\bea
{\cal L}\(a,\frac{\dot{a}}{N}\) \; = \; \F\(H\) \, .
\eea
A Hamiltonian analysis of the action \eqref{FLRWaction} with ${\cal L} = \F(H)$ clarifies the link with LQC. Due to the invariance under time reparametrization, the lapse $N$ is still a Lagrange multiplier and the only non-trivial pairs of canonically conjugate pairs of variables are
\bea
\{a,\pi_a\} \, = \, 1 \, =\, \{\psi,\pi_\psi\} \, .
\eea
The Lagrangian is clearly not degenerate and the momenta are given in terms of the velocities by
\bea\label{eqforp}
\pi_a \, = a^2\left[\, -\frac{3 H}{4 \pi G} + a^2 \F'\!(H) \right]\, , 
\qquad \pi_\psi = \f{a^3}{N} \dot{\psi} \, ,
\eea
where $\F'$ is the derivative of the function $\F$. The shape of the LQC effective Hamiltonian suggests the ansatz
\bea\label{ansatz}
\pi_a \, = \, \alpha \, a^n \arcsin \(\beta \, \frac{\dot{a}}{Na}\) \, ,
\eea
where $n$, $\alpha$ and $\beta$ are constants to be fixed. The condition that the momentum $\pi_a$ should be approximately given by the classical result $\pi_a = - 3 a \dot{a} / 4 \pi G N$ at low curvatures (or small $H$) sets $n=2$ and further requires that $\alpha \beta = -3 / 4 \pi G$.  Then, for this ansatz \eqref{ansatz} for $\pi_a$, $\F$ must be
\bea
\F(H) \, = \, \alpha \, H \arcsin(\beta H) + \frac{\alpha}{\beta} \sqrt{1-\beta^2 H^2} + \f{3 H^2}{8 \pi G} 
-\frac{\alpha}{\beta} 
\, ,
\eea
where the  integration constant has been fixed so that $\F(0)=0$, which means that one recovers the standard general relativity action in the low curvature regime.

Given this explicit form for the Lagrangian, one finds for the Hamiltonian density
\bea \label{st-ham}
{\cal H} \, = \,a^3 \left(  \frac{\pi_\psi^2}{2a^6} - \frac{8\pi G}{3}\alpha^2 \sin^2\(\frac{\pi_a}{2\alpha a^2}\) \right) \, .
\eea

Using the same procedure as in the previous subsection, one easily finds that the modified Friedman equation is given by
\bea
H^2 \, = \, \frac{8\pi G}{3} \, \rho \left(1 - \frac{\rho}{\rho_c}\right) \quad \text{with} \quad
\rho=\frac{\pi_\psi^2}{2a^6}\, \quad \text{and} \quad \rho_c=\frac{8\pi G}{3}\alpha^2 \,.
\eea
This coincides with the LQC effective dynamics provided
\bea
\alpha \; = \; \  \frac{3}{ 8\pi G \gamma \sqrt{\Delta}}.
\eea
Hence, $\alpha$ is determined by Newton's constant and the Barbero-Immirzi parameter. We recover the result of Chamseddine and Mukhanov from a Hamiltonian point of view. Let us emphasize that exactly the same function has been found in a rather different context much earlier in \cite{Date:2008gq, Helling:2009ia} from a Lagrangian point of view.

Before showing the large class of scalar-tensor theories whose Hamiltonian constraint reduces to \eqref{st-ham} for spatially flat FLRW space-times, we will show that this result can be extended to allow for non-vanishing spatial curvature.

\subsection{Spatial Curvature}
\label{subsec:curv}

It is possible to  generalize the previous procedure to the case of  a spatially curved FLRW space-time. In classical general relativity, one gets an additional contribution from the Einstein-Hilbert term coming due to  the 3-dimensional curvature ${}^3 \! R$  evaluated in a non-flat homogeneous and isotropic space-time. Hence, we start with the action 
\bea \label{action-curv}
S_k[a,N,\psi] \; = \; \int dt \left( - \frac{3a \dot{a}^2}{8 \pi G N} + a^3 \frac{\dot{\psi}^2}{N} + \f{3 N k a}{8 \pi G} + Na^3 {\cal L}_k\(a,\frac{\dot{a}}{N}\)\right)\, ,
\eea
where $k$ denotes the usual spatial curvature parameter. As in previous subsection, the extra term ${\cal L}_k$ (such that ${\cal L}_0={\cal L}$) contains the (yet unknown) additional terms that are necessary  to obtain the LQC effective dynamics in place of the classical general relativity ones. 

\subsubsection{Curved Cosmology in Mimetic Gravity}

We now assume that the curvature dependence can be taken into account by simply adding to ${\cal L}$ obtained in the flat case a new term that depends on the scale factor but not on its derivatives:
\bea \label{L-curved}
{\cal L}_k\(a,\frac{\dot{a}}{N}\) \; = \; \F\(H\) \, - \, \f{3 }{8 \pi G} \,V_k(a) \,  .
\eea
The potential-like term $V_k$ must vanish for $k=0$, in order to recover the results of the flat case. The overall normalization is chosen for later convenience. 

Since the two new terms in the action \eqref{action-curv} depend only on $a$ (and not on $\dot a$), it is straightforward to generalize the Hamiltonian analysis of the flat case.  The new Hamiltonian constraint is given by
\bea \label{ham-curv}
{\cal H} \, = \, a^3 \left(  \rho - \rho_c \sin^2\(\frac{\pi_a}{2\alpha a^2}\) - \frac{3k}{8 \pi G a^2} + \f{3 V_k(a)}{8 \pi G} \right) \, ,\quad
\rho=\frac{\pi_\psi^2}{2a^6},
\eea
and then the modified Friedmann equation becomes
\bea \label{f-curv}
H^2 \, = \, \left( \frac{8\pi G}{3} \rho - \frac{k}{a^2} + V_k(a) \right) 
\left(1 - \frac{1}{\rho_c} \left[{\rho} - \frac{3k}{8 \pi G a^2} + \f{3 V_k(a)}{8 \pi G} \right] \right) \,.
\eea
One can now compare this modified Friedmann equation, based on the very simple ansatz \eqref{L-curved}, with 
the LQC effective Friedmann equation obtained for spatially curved FLRW space-times.

\subsubsection{Curved Cosmology in Effective LQC and Quantization Ambiguities}
\label{ssubsec:curv-lqc}

Before making this comparison, it is important to review a quantization ambiguity which arises when performing the loop quantization of a homogeneous space-time with non-vanishing spatial curvature: this quantization ambiguity concerns the precise quantity that is to be expressed in terms of Planck-length holonomies.  Due to this quantization ambiguity, there exist three quantization prescriptions that give slightly different effective theories, which can each be compared to \eqref{f-curv}.

To be more specific concerning this quantization ambiguity, for closed FLRW space-times there exists a direct generalization of the procedure followed for the spatially flat space-time reviewed above in Sec.~\ref{ssubsec:lqc-q}, i.e., to express the field strength in terms of the holonomy of the Ashtekar-Barbero connection around a closed loop \cite{Ashtekar:2006es, Szulc:2006ep}.  This is known as the `F' loop quantization.  The `F' loop quantization is not possible for the open FLRW space-time, nor for spatially curved Bianchi space-times (the problem is that, when expressing the field strength in this fashion, the resulting function is not almost-periodic in the connection, and so cannot be promoted to be an operator in the quantum theory, see \cite{Vandersloot:2006ws, Ashtekar:2009um, WilsonEwing:2010rh} for details).  For the Bianchi space-times, an alternative way forward is to express the connection itself (rather than the field strength) in terms of a Planck-length holonomy, as proposed in \cite{Ashtekar:2009um, WilsonEwing:2010rh}; this alternative quantization is known as the `A' loop quantization.  However, for the open FLRW space-time, neither the `F' nor the `A' loop quantizations are viable.  Instead, it has been proposed that the `K' quantization, where one considers `holonomies' of the extrinsic curvature rather than of the Ashtekar-Barbero connection, may provide a reasonable approximation to a proper loop quantization \cite{Vandersloot:2006ws}.  Interestingly, the `F', `A', and `K' loop quantizations are all possible in the closed FLRW model, and it is possible to compare the resulting effective theories that result from each of these quantization prescriptions \cite{Corichi:2011pg, Singh:2013ava}, with the surprising result that the `K' quantization in fact appears to provide a better approximation to the `F' loop quantization than the `A' loop quantization does.  (All three quantizations are also possible for the spatially flat FLRW space-time and also Bianchi type I models, but in both cases all three quantizations turn out to be exactly equivalent.)  Therefore, for the closed FLRW space-time it is possible to compare the effective theories resulting from the `F', `A' and `K' quantization prescriptions to \eqref{f-curv}, while for the open space-time only the effective theory resulting from the `K' quantization is known.

Let us start with the `F' loop quantization.  In this case, the LQC effective Friedmann equation is \cite{Ashtekar:2006es, Corichi:2011pg}
\begin{align}
H^2 =& \( \f{8 \pi G}{3} \rho - \f{k}{a^2} - \f{k}{\gamma^2 a^2} + \f{1}{\Delta \gamma^2} \sin^2 \f{\sqrt{k \Delta}}{a} \) \nn \\ & \times \( 1 - \f{1}{\rho_c} \[\rho - \f{3k}{8 \pi G a^2} - \f{3k}{8 \pi G \gamma^2 a^2} + \f{3}{8 \pi G \Delta \gamma^2} \sin^2 \f{\sqrt{k\Delta}}{a} \] \),
\end{align}
which is precisely of the form of \eqref{f-curv} with
\be\label{Vk}
V_k(a) = - \f{k}{\gamma^2 a^2} + \f{k}{\Delta \gamma^2} \sin^2 \f{\sqrt{k\Delta}}{a}.
\ee
Note that it is not guaranteed that the effective Friedmann equation has the general form of \eqref{f-curv} for some $V_k(a)$, and therefore this result is encouraging because it indicates that the same modified gravity theory can be used to describe the LQC effective dynamics of both spatially flat and closed FLRW space-times (assuming the `F' quantization for the closed FLRW space-time).

Now, while it is not known how to perform a proper loop quantization for the open FLRW space-time, it appears reasonable to assume that the effective dynamics for such a loop quantization would also have the form \eqref{f-curv} where $V_k(a)$ is  \eqref{Vk} with the only difference that now $k$ is negative.  If this is indeed the case, then the resulting LQC effective Friedmann equation for the open FLRW space-time would be
\begin{align}
H^2 =& \( \f{8 \pi G}{3} \rho + \f{|k|}{a^2} + \f{|k|}{\gamma^2 a^2} - \f{1}{\Delta \gamma^2} \sinh^2 \f{\sqrt{|k| \Delta}}{a} \) \nn \\ & \times 
\( 1 - \f{1}{\rho_c} \[\rho + \f{3|k|}{8 \pi G a^2} + \f{3|k|}{8 \pi G \gamma^2 a^2} - \f{3}{8 \pi G \Delta \gamma^2} \sinh^2 \f{\sqrt{|k|\Delta}}{a} \] \).
\end{align}

Moving on to the `A' loop quantization, the effective equation for the closed FLRW space-time turns out to have a form which is different  than \eqref{f-curv} \cite{Corichi:2011pg}.  This provides an explicit example that shows that, in general, generic modifications to the Friedmann equations cannot be written in the form \eqref{f-curv}.  It is only in some special cases that it will be possible to describe quantum gravity effects via a mimetic scalar-tensor theory, as is the case for the `F' loop quantization.

Finally, for the `K' loop quantization the effective Friedmann equation for the closed and open FLRW space-times is \cite{Vandersloot:2006ws, Singh:2013ava}
\be \label{lqc-k}
H^2 = \( \f{8 \pi G}{3} \rho + \f{k}{a^2} \) \times \( 1 - \f{1}{\rho_c} \[\rho + \f{3k}{8 \pi G a^2} \] \),
\ee
where $k>0$ for a closed universe and $k<0$ for an open universe.  Interestingly, in this case we find that the effective Friedmann equation is again of the form \eqref{f-curv}, this time with $V_k(a) = 0$.  There are three important points here.  First, for the `K' quantization, which can be performed for both open and closed FLRW space-times, the effective theory can be understood to come from a scalar-tensor theory, and the same theory describes equally well space-times with positive or negative spatial curvature.  Second, in agreement with earlier results \cite{Corichi:2011pg, Singh:2013ava}, we find that the `K' quantization is more similar to the `F' quantization than the `A' quantization is, in that its resulting effective Friedmann equations can be described by a scalar-tensor theory.  And third, the `K' quantization does not require a $V_k(a)$ term in the action \eqref{action-curv}.  Since a non-vanishing $V_k(a)$ term breaks Lorentz invariance (as we shall discuss in more detail in the following section), it is quite interesting that the effective theory of the `K' quantization in fact corresponds to a modified gravity theory which is Lorentz invariant.

To summarize, there is a quantization ambiguity that arises in LQC when considering spatially curved homogeneous space-times.  There exist `F', `A' and `K' quantizations for the closed FLRW space-time, while only the `K' quantization is known for the open topology.  For closed FLRW space-time, the effective equations resulting from the `F' quantization prescription (but not the `A' quantization) can be understood to come from a mimetic gravity theory with the action \eqref{action-curv}, for a particular choice of $V_k(a)$.  This mimetic gravity theory in fact provides a candidate effective theory for a proper loop quantization of the open FLRW space-time.  Finally, the effective equations for the `K' quantization of FLRW space-times (for all $k$) can also be understood to follow from a mimetic gravity action of the form \eqref{action-curv}, in this case with $V_k(a) = 0$.

In short, these results show that the LQC effective equations (with the exception of the `A' quantization) for homogeneous and isotropic cosmologies are identical to the Friedmann equations of a particular mimetic gravity theory.

\section{Effective Loop Quantum Gravity and Mimetic Gravity}
\label{sec:lqg}

In this section, we address the question of finding a covariant action of modified gravity that reduces to
\bea\label{cosmoaction}
S[a,N,\psi] = \int dt  Na^3\left( \frac{\dot{\psi}^2}{2 N^2} - \f{\rho_c}{2}  \left[ \beta H \arcsin(\beta H) + \sqrt{1-\beta^2 H^2} - 1\right]\right) \, , \quad 
\beta^2 = \f{3}{2 \pi G \rho_c} \, ,
\eea
in a spatially flat homogeneous and isotropic space-time (the non-flat case will be discussed below), as found in the previous section. Of course, such a condition is not very restrictive and one can expect that many different covariant actions could yield the same cosmological action. 

Because of the non-linearity of the Lagrangian in the Hubble parameter $H$, it is not possible to find an $f(R)$ theory which exactly reproduces \eqref{cosmoaction} in the cosmological sector, although an approximate construction has been found in \cite{Olmo:2008nf} (where one considers $f(R)$ theories \`a la Palatini). The reason is that the Ricci scalar, given by 
\bea
R \; = \; 6 \left( \frac{1}{a} \frac{1}{N} \frac{d}{dt} \left( \frac{\dot{a}}{N}\right) + \left(\frac{\dot{a}}{Na}\right)^2\right)\, ,
\eea
involves second derivatives of the scale factor and this prevents us from recovering a Lagrangian $f(R)$ 
that depends only on $a$ and $\dot{a}$ in the cosmological sector. 

Another possibility would be to consider an action involving nonlinear combinations of the Riemann tensor.  It was shown in \cite{Helling:2009ia} that a Lagrangian containing contractions of the Ricci tensor of the form $R_{\mu_1}{}^{\mu_2} R_{\mu_2}{}^{\mu_3} \cdots R_{\mu_n}{}^{\mu_1}$ indeed reduces to \eqref{cosmoaction} for spatially flat FLRW space-times.  However, the explicit expression of this Lagrangian is very cumbersome and it does not appear to be suited for calculations away from the homogeneous and isotropic sector. Furthermore, as the Lagrangian involves higher powers of the Ricci tensor, the theory will propagate more degrees of freedom than the usual two tensor modes, and these additional degrees of freedom will lead to Ostrogradsky instabilities. 

Following \cite{Chamseddine:2016uef},  we will try to explore scalar-tensor actions, in particular mimetic theories. Following the results recalled in the previous sections, we look for a covariant action of the form
\bea\label{actionV}
S[{g}_{\mu\nu},\phi] \, = \, \int d^4x \, \sqrt{-g} \left( {\f{f {R}}{16 \pi G}} + {\cal L}_\phi(\phi,\phi_\mu, \phi_{\mu\nu})
+ \lambda ({X}+1) - \frac{1}{2}{g}^{\mu\nu} \psi_\mu \psi_\nu \right)
\eea
where $f$ is a function of $\phi$ only and ${\cal L}_\phi$ is a scalar function which depends on $\phi$ and its first and second derivatives  $\phi_\mu$ and $\phi_{\mu\nu}$ only. 
 
As we said above, the solution of our problem is far from being unique and one can find an infinite class of solutions with few restrictions on the Lagrangian ${\cal L}_\phi$. Here, we want to propose the simplest class of solutions which generalize the Chamseddine-Mukhanov model, and we restrict ourselves to higher-derivative Lagrangians for the scalar field of the form
\bea
{\cal L}_\phi \; = \; {\cal L}_\phi\(\phi, L^{(2)}_1,L^{(2)}_2,L^{(2)}_3,L^{(2)}_4,L^{(2)}_5\)
\eea
where the $L^{(2)}_A$ are the five elementary quadratic Lagrangians introduced in \eqref{QuadraticL}. Of course, one could also consider further generalizations including the cubic elementary Lagrangians. 
	
Now, we find the conditions that $f$ and ${\cal L}_\phi$ must satisfy for the action \eqref{actionV} to reduce to \eqref{FLRWaction} when the metric ${g}_{\mu\nu}$  is \eqref{FLRWmetric} and the fields $\phi$ and $\psi$ depend on time only. In this case,
\bea \label{L-cosmo1}
L^{(2)}_1 & = & \left(\Dt^2 \phi\right)^2 + 3 \left(\Dt\phi \, \frac{\Dt a}{a}\right)^2 \, ,\quad
L^{(2)}_2  =  \left[\frac{1}{a^3} \Dt \left( a^3\,  \Dt\phi \right)\right]^2 \, ,\\
L^{(2)}_3 & = & -\frac{1}{a^3} \left(\Dt\phi\right)^2  
\Dt^2\phi \ \Dt( a^3 \Dt\phi) \, ,\\
L^{(2)}_4 & = & -  \left(\Dt\phi \, \Dt^2\phi   \right)^2 \, ,\quad
L^{(2)}_5  =  \left[\left(\Dt\phi\right)^2   \Dt^2\phi\right]^2 \, , \label{L-cosmo5}
\eea
where we have introduced the notation
\be
\Dt\varphi\equiv \frac{\dot \varphi}{N}
\ee
for any time-dependent function $\varphi(t)$.

The advantage to consider mimetic theories is that the time derivative of the scalar field is automatically normalized since we have, in the cosmological background, 
\bea
X\;=\;- \left(\frac{\dot{\phi}}{N}\right)^2 \; = \; -1\,.
\eea
As consequence, in the cosmological context, the elementary quadratic Lagrangians simplify drastically:
\bea
L^{(2)}_1= 3 \left(\frac{\dot{a}}{Na}\right)^2 , \; L^{(2)}_2= 9 \left(\frac{\dot{a}}{Na}\right)^2 ,\;  L^{(2)}_3=L^{(2)}_4=L^{(2)}_5=0 \; .
\eea
Due to the form of \eqref{FLRWaction}, the functions $f$ and ${\cal L}_\phi$  are necessarily independent of $\phi$. All these ingredients allow us to conclude immediately that the action \eqref{actionV} reproduces the LQC effective dynamics when the Lagrangian ${\cal L}_\phi$ takes the form
\bea \label{solflat}
{{\cal L}_\phi(L^{(2)}_1,L^{(2)}_2,L^{(2)}_3,L^{(2)}_4,L^{(2)}_5) \; = \; \( \sum_{i=1}^5 \alpha_i L^{(2)}_i \) + 
{U}(L^{(2)}_1,L^{(2)}_2,L^{(2)}_3,L^{(2)}_4,L^{(2)}_5)}
\eea
with the conditions
\bea \label{cond}
{\alpha_1 + 3 \alpha_2 = \frac{f}{8 \pi G} \, , \quad
{U}(3H^2,9H^2,0,0,0) = \f{\rho_c}{2} \left(1 - \sqrt{1-\beta^2 H^2} - \beta H \arcsin(\beta H)  \right)\, ,}
\eea
for any value of $H$. The mimetic gravity model proposed by Chamseddine and Mukhanov clearly belongs to this class and consists in taking $f = 24 \pi G \alpha_2 = 1$, with all other $\alpha_A=0$, and ${U}$ is assumed to be a function of $L^{(2)}_2=(\Box\phi)^2$ only. However, this result shows that there exist different possibilities to get the LQC effective dynamics. Nonetheless, in a theory of mimetic gravity with the 
condition $X+1=0$, the Lagrangians $L_3^{(2)}$, $L_4^{(2)}$ and $L_5^{(2)}$ are always vanishing, not only in the cosmological sector.
This is an immediate consequence of the fact that $\phi_{\mu\nu}\phi^\mu=0$.  Thus, one can restrict the function ${\cal L}_\phi$ to depend
only on the first two arguments without loss of generality. 

\medskip

Let us close this section with a discussion on the case of a non-flat cosmology. The only difference with what has been done so far is that the scalar-tensor action, when evaluated for a spatially curved FLRW space-time, must include the additional $V_k(a)$ term that arises in \eqref{L-curved} (at least for the `F' loop quantization, but not for the `K' quantization, as explained in \ref{ssubsec:curv-lqc}). Such a term can be obtained by adding to the effective covariant action \eqref{actionV} with ${\cal L}_\phi$ given by \eqref{solflat} a new potential term
\bea
-\int d^4x \sqrt{\vert g \vert} \, {\cal V}(g_{\mu\nu},\phi)
\eea
which reproduces exactly $-V_k(a)$ when evaluated on a curved FLRW space-time. As in the previous situation, we obviously do not expect the solution for ${\cal V}$ to be unique. However, we can exhibit some properties $\cal V$ must have. The fact that $V_k(a)$ is a highly non-linear function (even non-polynomial) of the scale factor $a$ which does not depend on its time derivatives implies that ${\cal V}$ cannot be constructed from the full space-time components of the Riemann tensor only (with no contractions with derivatives of $\phi$) nor from the elementary scalar-tensor Lagrangians $L_A^{(2)}$. The reason is that, when evaluated on a curved FLRW space-time, such terms necessarily produce derivatives of the scale factor which cannot be removed by integration by parts. Cubic (or even higher-order) scalar-tensor terms as $L_A^{(3)}$ produce the same problem. Only the components of the 3-dimensional Riemann tensor are functions of $a$ only when evaluated on a curved FLRW solution. This can be easily illustrated with the 3-dimensional Ricci scalar ${}^3\! R$ which reduces to 
\bea
{}^3\! R \; = \;  - \frac{6k}{a^2} \, .
\eea
From this point of view, the potential ${\cal V}(g_{\mu\nu},\phi)$ can be constructed from the 3-dimensional Riemann tensor, in which case it breaks the Lorentz invariance. Furthermore, as $V_k$ can be expanded as the following series
\bea
V_k(a) \; = \; \sum_{n>0} v_n \left(\frac{6k}{a^2}\right)^n \; \equiv \; \tilde{V}(-\frac{6k}{a^2})\, ,
\eea
one can choose for the potential ${\cal V}_k$ a function which depends only on ${}^3\!R$ according to
\bea
{\cal V}(g_{\mu\nu},\phi) \; = \; \tilde{V}({}^3\! R ) \, .
\eea
Interestingly, similar potential are considered in Ho{\v{r}}ava-Lifshitz-type models of gravity to construct renormalizable theories of gravity. Note that, using the St\"uckelberg method, it is nonetheless possible to render ${\cal V}$ fully covariant using the scalar field $\phi$ as a covariant clock. The price to pay is that the covariant version of ${\cal V}$ would involve terms like $R_{\mu\nu\rho\sigma} \phi^\mu \phi^\nu \phi^\rho \phi^\sigma$ in the action which would most probably introduce many unhealthy degrees of freedom in the theory.

\section{Perspectives}
\label{sec:perspectives}

In this paper, we have constructed a family of higher-derivative scalar-tensor theories that possess the property to reproduce exactly the effective dynamics of loop quantum cosmology for flat, closed and open homogeneous and isotropic space-times, leading to bouncing solutions. This family thus generalizes the particular model considered by Chamseddine and Mukhanov for the spatially flat case \cite{Chamseddine:2016uef}.

An important question is whether this family, identified only at the level of the homogeneous and isotropic dynamics, contains a theory that could fully represent an effective description of the full  loop quantum gravity (LQG). Indeed, a study limited to homogeneous cosmologies is too restrictive to be conclusive and one should  go beyond homogenous and isotropic solutions to test further the interest of the theories we have identified.
In particular, it would interesting to investigate whether some of them can adequately describe anisotropic space-times, cosmological perturbations, black holes, and more. 

Concerning anisotropic space-times, it should be stressed  that   the equations of motion for the Bianchi I space-time  derived  in the specific mimetic theory of \cite{Chamseddine:2016uef} do not coincide with  the LQC effective equations given in \cite{Chiou:2007sp, Ashtekar:2009vc}, even if their qualitative dynamics are quite similar (see, e.g., \cite{Gupt:2012vi}).  It would thus be worth  studying whether another theory among the family identified in the present work is able to reproduce these Bianchi I equations. 
 We leave this question for future work.
 
It would also be interesting to extend our study to include linear cosmological perturbations around the FLRW background.  Indeed, in the presence of perturbative inhomogeneities a number of important qualitative similarities have already been noticed between some modified gravity theories and the LQC effective dynamics \cite{Cai:2014zga}. One might hope that there is in fact an exact correspondence for at least one  mimetic theory of gravity. Another point is that, as shown in \cite{Firouzjahi:2017txv}, ghost and gradient instabilities arise when considering cosmological perturbations for the action of the mimetic theory proposed in \cite{Chamseddine:2016uef} (note that these instabilities are different from the higher derivative ghosts which mimetic gravity is safe from as a DHOST theory); it would be interesting to see if this is the case for all actions in the mimetic family, or if there exist some mimetic actions without ghost or gradient instabilities for cosmological perturbations.

Furthermore, it would be interesting to study black hole solutions for the scalar-tensor theories  considered here.  For the specific mimetic theory proposed in \cite{Chamseddine:2016uef}, the spherically symmetric black hole solution has been studied in \cite{Chamseddine:2016ktu}. We now have at our disposal  a  large class of mimetic  theories to consider.  In particular, it might be possible to find solutions related to those constructed in the context of LQG such as the Planck stars \cite{Rovelli:2014cta, DeLorenzo:2014pta}.

Beyond the natural extensions discussed above, it should also be mentioned that an  alternative description of the LQC effective dynamics offers a complementary framework that could provide new insights.  One such example is the question of a possible signature change in LQC: due to a modification in the Dirac algebra of the (effective) constraints of LQC, it has been suggested that the signature of the metric may change from Lorentzian to Euclidean around the bounce point, see \cite{Cailleteau:2011kr} for details.  While this question is difficult to address in LQC due to the gauge-fixings that are necessary before quantization, the metric clearly remains Lorentzian at all times in the mimetic theories considered here and  this could suggest that there is no signature change in LQC.

Finally, this result suggests some new links between LQG, Ho{\v{r}}ava-Lifshitz gravity, and non-commutative geometry.  As seen in Sec.~\ref{ssubsec:curv-lqc}, for some (but not all) versions of LQC it is necessary to add a Lorentz-violating term in order to recover the correct effective Friedmann equations in the presence of scalar curvature, along the lines of what is done in Ho{\v{r}}ava-Lifshitz gravity.  As Ho{\v{r}}ava-Lifshitz gravity is known to be perturbatively renormalizable, it is possible that this effective theory may be renormalizable as well.  In addition, the mimetic condition \eqref{mimcondition} has been argued to arise naturally in non-commutative geometry when requiring the quantization of the three-dimensional volume \cite{Chamseddine:2014nxa}.  It is intriguing that it is precisely modified gravity theories that satisfy this condition which give the LQC effective dynamics for isotropic cosmologies.  An exploration in greater depth of the links between these different approaches to the problem of quantum gravity may provide important new insights.

\bigskip

\noindent
\emph{Note:} During the completion of  this paper, we were informed that the independent work \cite{Bodendorfer:2017} also points out the relation between the mimetic gravity theory of \cite{Chamseddine:2016uef} and the effective equations of LQC.

\acknowledgments

We would like to warmly thank Norbert Bodendorfer for valuable discussions and for telling us of his work \cite{Bodendorfer:2017} before its publication.

\raggedright

\end{document}